\documentclass[twocolumn]{aastex62}

\received{XXXX XX, 201X}
\revised{XXXX XX, 201X}
\accepted{\today}
\submitjournal{ApJ}

\shorttitle{Modelling of the quasar main sequence}
\shortauthors{Panda et al.}
\begin{document}

%\pagecolor{black}
%\color{white}

\title{Modelling of the quasar main sequence in the optical plane}

\correspondingauthor{Swayamtrupta Panda}
\email{spanda@camk.edu.pl}

\author{Swayamtrupta Panda}
\affiliation{Center for Theoretical Physics (PAN), Al. Lotnik{\'o}w 32/46, 02-668 Warsaw, Poland}
\affiliation{Nicolaus Copernicus Astronomical Center (PAN), ul. Bartycka 18, 00-716 Warsaw, Poland}

\author{Bo{\.z}ena Czerny}
\affiliation{Center for Theoretical Physics (PAN), Al. Lotnik{\'o}w 32/46, 02-668 Warsaw, Poland}
\affiliation{Nicolaus Copernicus Astronomical Center (PAN), ul. Bartycka 18, 00-716 Warsaw, Poland}

\author{Tek P. Adhikari}
\affiliation{Nicolaus Copernicus Astronomical Center (PAN), ul. Bartycka 18, 00-716 Warsaw, Poland}

\author{Krzysztof Hryniewicz}
\affiliation{Nicolaus Copernicus Astronomical Center (PAN), ul. Bartycka 18, 00-716 Warsaw, Poland}

\author{Conor Wildy}
\affiliation{Center for Theoretical Physics (PAN), Al. Lotnik{\'o}w 32/46, 02-668 Warsaw, Poland}

\author{Joanna Kuraszkiewicz}
\affiliation{Harvard-Smithsonian Center for Astrophysics, Cambridge, MA 02138, USA}

\author{Marzena {\'S}niegowska}
\affiliation{Center for Theoretical Physics (PAN), Al. Lotnik{\'o}w 32/46, 02-668 Warsaw, Poland}
\affiliation{Warsaw University Observatory, Al. Ujazdowskie
 4, 00-478 Warsaw, Poland}



\begin{abstract}
The concept of the quasar main sequence is very attractive since it stresses correlations between various parameters and implies the underlying simplicity. In the optical plane defined by the width of the H$\beta$ line and the ratio of the equivalent width of the Fe II to H$\beta$ observed objects form a characteristic pattern. In this paper, we use a physically motivated model to explain the distribution of quasars in the optical plane. Continuum is modelled as an accretion disk with a hard X-ray power law uniquely tight to the disk at the basis of observational scaling, and the Broad Line Region distance is determined also from observational scaling. We perform the computations of the FeII and H$\beta$ line production with the code CLOUDY. We have only six free parameters for an individual source: maximum temperature of the accretion disk, Eddington ratio, cloud density, cloud column density, microturbulence, and iron abundance, and only the last four remain as global parameters in our modelling of the whole sequence. Our theoretically computed points cover well the optical plane part populated with the observed quasars, particularly if we allow for super-Solar abundance of heavy elements. Explanation of the exceptionally strong Fe II emitter requires a stronger contribution from the dark sides of the clouds. Analyzing the way how our model covers the optical plane we conclude that there is no single simple driver behind the sequence, as neither the Eddington ratio nor broad band spectrum shape plays the dominant role. Also, the role of the viewing angle in providing the dispersion of the quasar main sequence is apparently not as strong as expected.
\end{abstract}




\section{Introduction} \label{sec:intro}

% I've made several changes to this section, mostly to improve the language and clarity

Quasars, or more generally, active galactic nuclei, are complex objects having a supermassive black hole located at the center of a galaxy accreting
matter. In high Eddington ratio (${\rm \lambda{}_{Edd}}$) sources, inflowing matter forms an accretion disk along with a surrounding medium in the form of a hot corona and a wind. In radio-loud (jetted AGN) we also see a collimated outflow. Further out, the Broad Line
Region (BLR) and the dusty/molecular torus are located, completing the picture. It is thus not surprising that active galaxies come in a
variety of types. If we concentrate on sources with a clear view of the nucleus (Type 1 sources), they
should display considerable variety, taking into account that the spectral morphology is affected by the black hole mass, accretion
rate, black hole spin, and the viewing angle. This last aspect is important even for Type 1 objects due to flattened geometry of the some elements (disk and BLR). Extinction and departures from the stationarity further complicate the picture.

The search for a Quasar Main Sequence has been undertaken in recent decades with surprising success. The study of \cite{bg92} used Principal Component Analysis (PCA) to show that a single parameter corresponding to Eigenvector 1 (EV1) is responsible for a significant fraction of the dispersion, 29.2\% of the total variance, in the studied quasar sample. In their analysis, EV1 represented a combination of 13 properties measured for each source, including several line equivalent widths, line shape parameters, absolute magnitude, and the broad band index $\alpha_{\mathrm{ox}}$. This line of study was further pursued by several authors \citep{dul96,sul00,bor02,kur09}. In recent studies, instead of a full EV1, a reduced EV1 was used, with four elements \citep{sul09,mar14,sul15}. The optical plane version was just based on two quantities: the ratio of the equivalent width (EW) of Fe II complex, measured in the 4434 - 4684 $\text{\AA}$ wavelength range to H$\beta$ range, $\mathrm{\mathrm{R_{Fe}}}$, and the Full Width at Half Maximum (FWHM) of H$\beta$ \citep{sh14}. The Quasar Main Sequence is thus well established, albeit it is not as narrow as the stellar main sequence in the Hertzsprung-Russell diagram \citep{sul00,sul01}.

% In the following paragraph I've noted that black hole mass is postulated to be behind the EV1 eddington ratio correlation in Shen and Ho

The key question is why such a sequence forms at all? Quasars are complex objects, \textcolor{blue}{their} central engine is described by the black hole mass, accretion rate (or Eddington ratio), black hole spin, and its properties are also affected by the viewing angle. Broad Line Region (BLR) itself is a complex extended region. It is not simple to identify just a single underlying parameters, although papers devoted to the quasar main sequence attempted to identify the actual driver of the sequence. The study of \cite{bg92} has postulated that the key parameter is the Eddington ratio, with \citep{sh14} tracing this to the black hole mass distribution among quasars. Additional dependence on extinction has also been reported \citep{kur09}. \citet{bonning07} mentioned the possibility that the key parameter is the maximum temperature of the accretion disk or, equivalently, the peak of the spectral energy density (SED) since the shape of the incident continuum should be responsible both for the continuum and emission line properties. Here we start from the generic approach: we model realistically the SED and the BLR setup, and we calculate the line emissivity using the code CLOUDY, version C17 \citep{f17}, and we compare them to the distribution of $\mathrm{\mathrm{R_{Fe}}}$ in high-quality data sub-sample of the objects studied by \cite{sh14}. As our leading parameters we use the maximum disk temperature and the Eddington ratio instead of the black hole mass and accretion rate with the aim to see if any of the previously proposed main sequence indeed plays a dominant role in the optical plane coverage.

\section{Model}
\label{sec:model}

% A few changes made here, nothing serious

We model the emission of the Fe II and H$\beta$ lines as functions of physically motivated parameters of an active nucleus. The problem of Fe II emission and its ratio to H$\beta$ has a long history. Fe II emission was first identified in the spectrum of the quasar 3C 273 by \cite{wo67}, and in Seyfert galaxy I Zw 1 by \cite{sar68}. \cite{oster76} studied in detail Mkn 376, a source with strong and broad Fe II emission. The width of the Balmer lines in this source is broad (FWHM of 5000 km/s). He identified the likely emission mechanism as the resonance fluorescence, and from the observed SED shape in this source estimated the Fe II to H$\beta$ ratio as $\sim 1$, but mentioned that the estimate might not be accurate since the Balmer decrement does not agree with photonization calculations.

The details of the Fe II excitation are still not fully known since modelling was not quite satisfactory (e.g., \citealt{col00}; \citealt{bal04}). The
importance of the hard X-rays, as proposed by \cite{dav79} is also not clear (e.g., \citealt{wil87}; \citealt{sha94};
\citealt{bor89}; \citealt{zheng90}; see also \citealt{wil99}). Microturbulence or additional collisional excitation seems to be required (e.g., \citealt{bal04}). The ratio of the Fe II opt. to H$\beta$ was successfully modeled by \cite{jol87} where high density (10$^{10} - 10^{12}\; \mathrm{cm^{-3}}$), high column density (10$^{23} - 10^{25} \;\mathrm{cm^{-2}}$) and low temperature clouds are assumed (6000 - 8000 K). They also stressed the need for shielding of the region from the observed power law continuum in order to have a low degree of ionization, which reduces H$\beta$. The ionization potential of hydrogen (13.5984 eV) and iron (7.9024 eV for Fe I, 16.1878 eV for Fe II) are very similar which makes modelling the broad range of the ratio of H$\beta$ and Fe II transitions difficult.   

In the present paper we use the current version of the code CLOUDY \citep{f17} which incorporates the required local processes. We assume that the emission of both Fe II and H$\beta$ comes from the same region. There are strong arguments that the emission regions of Fe II and H$\beta$ are actually related. Both Fe II and H$\beta$ belong to Low Ionization Lines (LIL) according to classification of \citet{collin88}. Both Fe II and H$\beta$ respond to the variations of the continuum. By analysing NGC 7603 spectra, \cite{kol00} found optical Fe II to vary with a similar amplitude to the Balmer lines. In \cite{ves05} it was concluded that in NGC 5548 Fe II varied with the amplitude of 50-75\% of that of H$\beta$, while they also attempted to measure the Fe II delays. Their estimate (few hundred days for Fe II) remained highly uncertain. In \cite{2008ApJ...673...69K} a measurement of the reverberation response time of 300 days for Fe II lines in Akn 120 implied an origin in a region several times farther away from the central source than H$\beta$. Similar results were reported in \cite{2013ApJ...773...24R}, \cite{2013ApJ...769..128B} and \cite{2014ApJ...783L..34C}. However, reliable reverberation of optical Fe II lines done by \citet{hu15} for a sample of nine AGNs found that the detected lags were comparable to H$\beta$. The FWHMs of both show significant correlation (\citealt{2016MNRAS.462.1256C}) although for high Eddington ratios a kinematic shift is visible (\citealt{hu08}), while others have found no such systematic redshift (\citealt{2012ApJ...752L...7S}).

\subsection{incident continuum}

\begin{figure*}
\centering
\includegraphics[height=4.5cm,width=18cm]{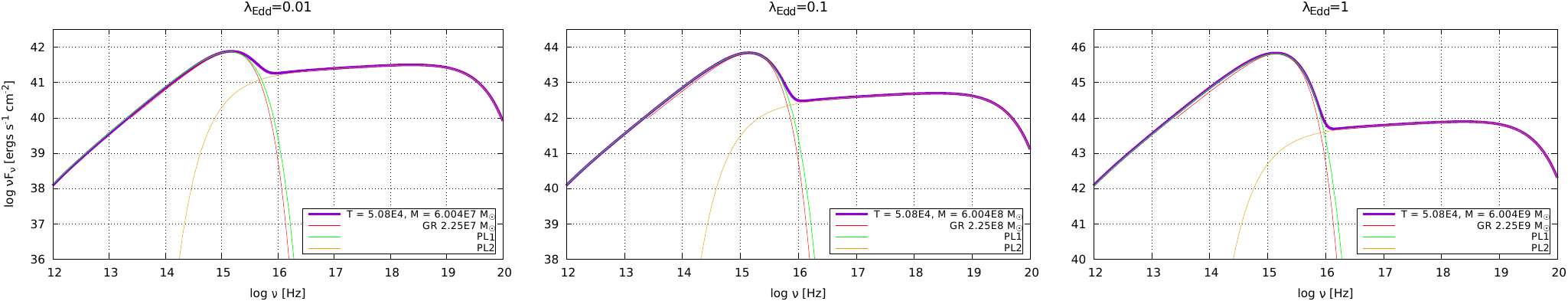}
\caption{Three examples of the incident continuum used in our computations: Big Blue Bump (green line: PL1) for $\mathrm{T_{max}} = 5.08\times 10^4\;\text{K}$ and three values of the Eddington ratio, which correspond to three values of the black hole mass (from left to right) $6 \times 10^7 M_{\odot}$, $6 \times 10^8 M_{\odot}$, and $6 \times 10^9 M_{\odot}$. The orange line (PL2) corresponds to the hard X-ray emission. The overall shape of the SED is given by the purple line. The red line corresponds to the classical model inclusive of the effects from GR.}
\label{fig:incident_continuum}
\end{figure*}

We parameterize the incident continuum as two power laws with low and high energy cutoffs. The optical/UV power law represents the emission from the accretion disk, where most of the energy is dissipated. This component forms the Big Blue Bump \citep{czerny87,richards06}. The slope of the power law $\alpha_{\mathrm{UV}}$ is assumed to be 1/3 (in convention $F_{\nu} \propto \nu^{\alpha}$, consistent with the theory of accretion disks \citep{ss73}, and supported by polarimetric observations of quasars \cite{ki08} and broad-band data fitting \citep{cap15}.  The high frequency cut-off, $\nu_*$ is the basic parameter which we vary in our model, i.e. we assume that the accretion disk luminosity can be described by the formula
\begin{equation}
  \mathrm{\nu L_{\nu} = A \nu^{4/3} exp (- \nu/\nu_*)}.
\label{eq:uv_power_law}
\end{equation}
This parameter is directly related to the maximum temperature in the accretion disk, $\mathrm{T_{max}}$.
If we use the Newtonian
 approximation of the disk from \citep{ss73}, there is a simple relation between the maximum temperature and global parameters of the accretion flow
\begin{equation}
\mathrm{T_{max}} = 1.732 \times 10^{19} \Big({\mathrm{\dot M \over M^2} \Big)^{1/4}\; [K]},
\label{eq:tmax}
\end{equation}
where the black hole mass, $M$, and the accretion rate, $\dot M$, are expressed in cgs units (see e.g. \citealt{panda_frontiers}). The disk maximum temperature is also related to the peak of the Spectral Energy Distribution (SED) of a Shakura-Sunyaev disk on the $\nu F_{\nu}$ plot
\begin{equation}
  {\mathrm{h\nu_{max}} \over  \mathrm{kT_{max}}} = 2.464. %2.092,
\end{equation}
Here $h$ is the Planck constant and $k$ is the Boltzmann constant. The simplified shape of the spectrum given by Eq.~\ref{eq:uv_power_law} peaks at the frequency $4 \nu_*/3$. Therefore, combining these two factor we finally obtain the convenient parameterization of the disk shape in the form
\begin{equation}
  \mathrm{\nu L_{\nu} = A \nu^{4/3} exp (-h\nu/1.853k\mathrm{T_{max}})}.
\label{eq:uv_power_law1}
\end{equation}

\begin{figure}[ht!]
  \centering
  \includegraphics[scale=0.4,angle=0]{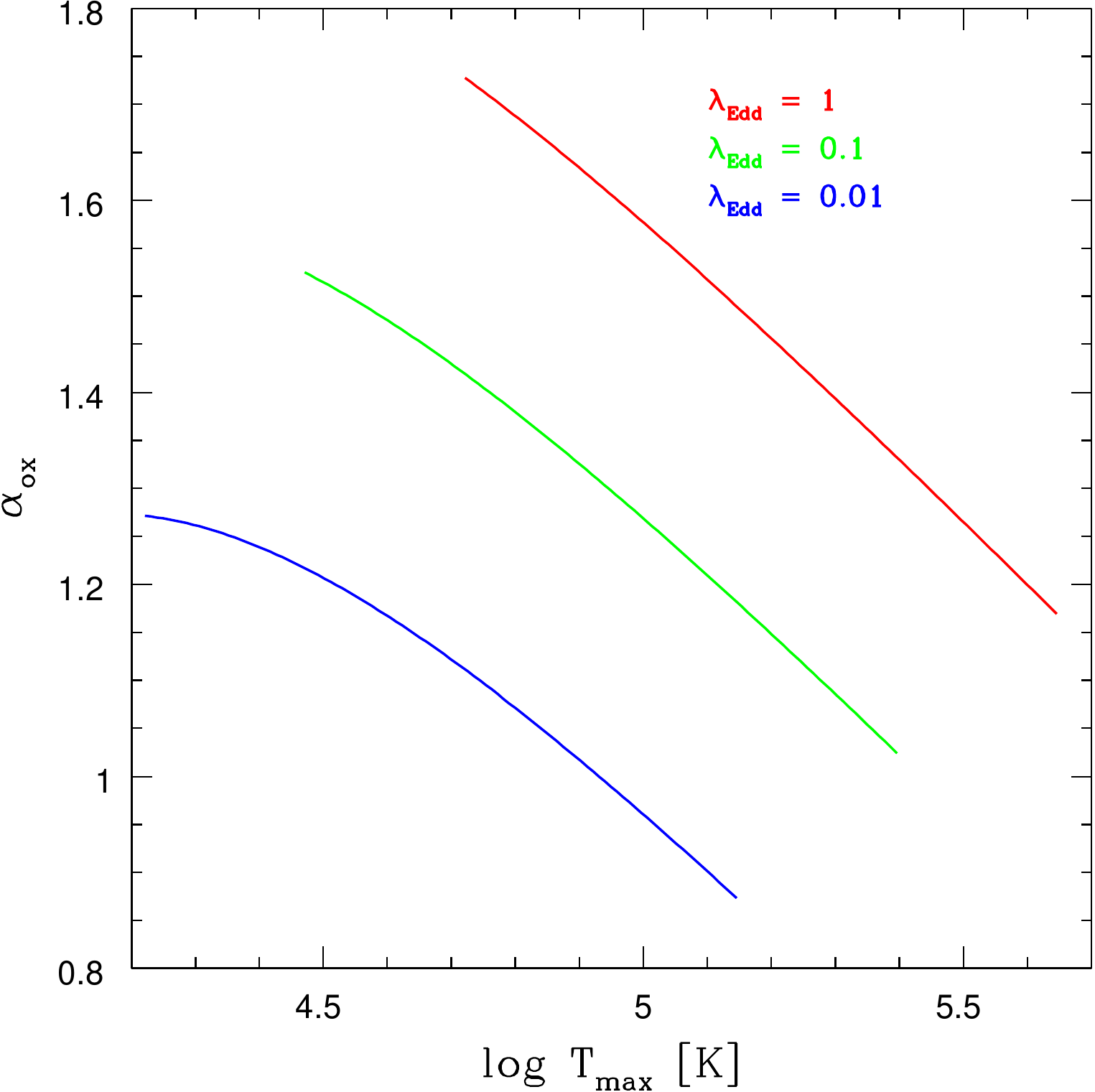}
  \caption{The range of the $\alpha_{\mathrm{ox}}$ index covered by our models. \label{fig:alphaox}}
\end{figure}

In the case of an accretion disk around a rotating black hole, the relativistic corrections are very important, the position of the Innermost Stable Circular Orbit strongly depend on spin, and the relations between the $\mathrm{T_{max}}$ and the global parameters of the accretion flow are more complex, and in that case more advanced back-modelling is necessary, as done by \citet{thomas16}. However, for the purpose of the current work we only consider non-rotating black holes which are well parameterized by the proposed prescription (see Fig.~\ref{fig:incident_continuum}), and are characterized by only two free parameters - black hole mass ($M$) and accretion rate (${\dot{M}}$).

The normalization of the optical/UV power law in Eq.~\ref{eq:uv_power_law} is given by the values of the black hole mass and accretion rate
\begin{equation}
 \mathrm{ A = 1.33 \times 10^{-20}  (\dot M M)^{2/3}\;[{\rm erg~s}^{-1} {\rm Hz}^{-4/3}]}
\end{equation}
under the assumption of the viewing angle $60^{\circ}$. Adopting such a viewing angle relates the integrated spectrum to the source bolometric luminosity. Small corrections due to on average smaller viewing angle of $\sim 40^{\circ}$, due to the obscuration by the dusty torus \citep{lawrence10}, is not an important factor.

AGN emit part of the energy in the form of the hard X-ray emission. We parameterize this component again as a power law with low and high energy cutoff. The high energy cutoff is set at 100 keV as in most fitting packages (e.g. OPTXAGN - standard AGN shape in CLOUDY, \citealt{done12}; \citealt{thomas16}). Observational data show a dispersion in this quantity from about 50 keV to over 200 keV but the measurements are still rare (for a compilation of measurements from NuSTAR, see \citealt{fabian15}).  

The relative normalization of the X-ray component with respect to UV is found from the universal scaling law recently discovered by \citet{lusso17}
\begin{equation}
%  \log L_X = 0.610 \log L_{UV} + 0.538 \log \mathrm{v_{FWHM}} + 6.00,
  \mathrm{\log L_X = 0.610 \log L_{UV} + 0.538 \log \mathrm{v_{FWHM}} + 3.40},
 \label{eq:Lusso_Risaliti}
\end{equation}
where $\rm{L_{UV}}$ is a monochromatic luminosity $\rm{\nu L_{\nu}}$ measured at 2500 \AA$\;$ and $\rm{L_X}$ is measured at 2 keV. The importance of the X-ray component decreases with the increase of the UV flux,  as illustrated in Fig.~\ref{fig:incident_continuum}. Equation \ref{eq:Lusso_Risaliti} gives us the value of the broad band index $\rm{\alpha_{ox}}$ measured between 2500 \AA~ and 2 keV. In Fig.~\ref{fig:alphaox} we show the range of this index covered in our computations. 

However, in order to use this formula we actually need the line Full Width at Half Maximum ($\mathrm{v_{FWHM}}$) which depends on the cloud location. We need this distance also for calculation of the line emissivity. 

\subsection{distance to the BLR}
\label{blr}
To determine the distance to the BLR clouds we use the observationally established relation from \citet{bentz13}. We choose the version {\it Clean} from their Table~14
\begin{equation}
  \mathrm{\log \mathrm{\mathrm{R_\mathrm{{BLR}}}} = 1.555 + 0.542 \log L_{44,5100}}~~~ [{\rm light~days}], 
\end{equation}
where the luminosity at 5100 \AA~is measured in units of $10^{44}$ egs s$^{-1}$.

Thus the shape of the incident radiation with its normalization provides us with the distance to the BLR. The value of the black hole mass gives the required line width
\begin{equation}
\rm{\mathrm{v_{FWHM}} = \bigg( {GM \over \mathrm{R_\mathrm{{BLR}}}}\bigg)^{1/2}},
\end{equation}
where we adopt the virial factor 1 for the BLR clouds. This is indeed a simplified approach since the virial factor seems to be a function of the measured line width but the average value is actually close to 1 \citep{meija18} which is enough for our current purpose.

Thus, for the two parameters characterizing the SED (e.g. normalization and the maximum disk temperature, or, equivalently, $M$ and $\dot M$) all the other parameters required to do the radiative transfer calculation are uniquely determined.

\subsection{radiative transfer}

We calculate H$\beta$ and optical Fe II emission performing the computations with the CLOUDY code \citep{f17}, version 17. We use the incident radiation and the distance to the BLR as described above. We adopt a traditional single cloud approximation \citep[e.g.][]{mushotzky84}. This is a reasonable approach since we are interested in a single line ratio, so going to the more complex Locally Optimized Cloud (LOC) cloud model of \citet{1995ApJ...455L.119B}  is not necessary. In addition, the BLR extension is not large, the outer to the inner ratio is estimated to be of order of 4 to 5 \citep{koshida14}. In CLOUDY computations we assume the plane parallel geometry which provides both emission from the illuminated side of the cloud and from the dark side of the cloud, in equal proportions. We assume that the density inside the cloud is constant. This is again an approximation since the clouds in the vicinity of the nucleus are expected to be in pressure equilibrium in order to survive at least in the dynamical timescale \citep[e.g.][]{rozanska06}. Some models based on radiation pressure confinement suggested that the density gradient in the cloud must be steep \citep{stern14,baskin18}. However, at least in a significant fraction of sources which show the presence of the Intermediate Line Region instead of two well separated BLR and NLR, the local density at the cloud surface must be high and the density gradient is rather shallow \citep{adhikari16,adhikari18}.
We assume two additional free parameters of the cloud: density, $n$, and hydrogen column density, $\mathrm{N_H}$.

H$\beta$ flux is taken from the code output, and Fe II emission is calculated by summing up all the Fe II lines in the range from 4434 \AA~to 4684 \AA, as defined in \citet{sh14}. The parameter $\mathrm{R_{Fe}}$ is calculated as the ratio of these two numbers.

\subsection{Summary of the model parameters}

For convenience, we decided to use the following parameters to present our results: $\mathrm{T_{max}}$, $L/L_\mathrm{{Edd}}$, $n$, and $\mathrm{N_H}$. Here $L/L_\mathrm{{Edd}}$ is defined using the Eddington accretion rate value $1.26 \times 10^{38} (M/M_{\odot})$, and to calculate the Eddington accretion rate we used the Newtonian accretion disk efficiency 1/12 \citep{ss73}. For the Eddington ratio, we choose values between 0.01 to 1. We cannot consider higher values since then the slim disk effects would be important \citep{abramowicz88,wang14}. At lower accretion rates inner optically thin flow likely replaces the cold geometrically thin optically thick disk, which leads to modification of both SED \citep{narayan94,nemmen14} and the BLR itself \citep[e.g.][]{czerny04,balmaverde15}. In general, we consider the range of the $\mathrm{T_{max}}$ between $1.6 \times 10^4$ K and $5 \times 10^5 $ K, appropriate for AGN Big Blue Bump. The corresponding range of the black hole masses depends on the adopted Eddington ratio. We give example values in Table 1. We assume that the viable range of masses is between $10^6 M_{\odot}$ and $3 \times 10^9 M_{\odot}$ for low redshift sources with detected H$\beta$ line. Thus not all of $\mathrm{T_{max}}$ range is realistic for a given choice of the Eddington ratio, and we include this effect in our plots.
%since it corresponds to the choice of the black hole mass range from $6.064\times 10^3\;M_{\odot}$ to $6.064\times 10^7\;M_{\odot}$ for $L/L_\mathrm{{Edd}} = 0.01$, and from $6.064\times 10^5\;M_{\odot}$ to $6.064\times 10^9\;M_{\odot}$ for $L/L_\mathrm{{Edd}} = 1.0$.
Therefore, the allowed range of the $\mathrm{T_{max}}$ for a given Eddington ratio is further constrained by the limits on the black hole mass to be between $10^6\;M_{\odot}$ and $5\times 10^9\;M_{\odot}$. 
For the cloud density $n$, we choose values from $10^{10}$ cm$^{-3}$ to $10^{12}$ cm$^{-3}$, appropriate for the LIL part of the BLR. The column density was assumed to vary from $10^{22}$ cm$^{-2}$ to $10^{24}$ cm$^{-2}$. We assume constant density cloud profile but we address the issue of the constant pressure clouds in the discussion. We also allow for the turbulent velocity since the need for a velocity of order of 10 - 20 km s$^{-1}$ is evident from the previous studies of the Fe II production \citep{bruhweiler08}.

Finally, AGN can have a range of metallicities, and most studies have found that the abundances are actually at least Solar and mostly super-Solar \citep[factor 1 to 10][]{hamann99,tortosa18}, even for high redshift quasars \citep[by a factor 5 to 10, e.g.][]{simon10}.
In our computations we either assumed the standard chemical composition which is the default in CLOUDY, or we allow for an increase in Fe abundance. In the first option the default values
are provided by CLOUDY, and in this case (see Table 7.1 of CLOUDY manual) C and O abundances come from
photospheric abundances of \cite{2001ApJ...556L..63A, 2002ApJ...573L.137A}, while N, Ne, Mg, Si, and Fe are from \cite{2001AIPC..598...23H}, and  the Fe is taken from \cite{2001AIPC..598...23H}.
 If we now specify a Solar abundance, Fe is taken from the GASS model (\citealt{2010Ap&SS.328..179G}), and in this option we also consider super-Solar abundance.

\begin{table}
  \caption{The examples of the correspondence between the black hole masses and the assumed $\mathrm{T_{max}}$ and $\lambda_\mathrm{{Edd}}$ (see Eq.\ref{eq:tmax}).}   % title of Table^M
  \label{tab:bhrange}      % is used to refer this table in the text^M
  \centering                          % used for centering table^M
  \begin{tabular}{l r r r r r r}        % centered columns (4 columns)^M
    \hline\hline      % inserts double horizontal lines^M
     $\lambda_\mathrm{{Edd}}$   & 0.01  &  0.01 & 0.1  &  0.1 & 1.0  & 1.0    \\
     $\log \mathrm{T_{max}}$   & 5.145  &  4.270 & 5.395  & 4.520 & 5.645 & 4.770    \\
    \hline
     $\log \mathrm{M/M_{\odot}}$   & 6.0  &  9.5 & 6.0  & 9.5 & 6.0  & 9.5    \\
  \end{tabular}
  \end{table}
  
%http://cdsads.u-strasbg.fr/abs/1976ApJ...203..329O  osterbrock76
%http://cdsads.u-strasbg.fr/abs/1987A%26A...184...33J joly1987
%Hu et al. (2015) http://cdsads.u-strasbg.fr/abs/2015ApJ...804..138H

\section{Results}
\label{sec:results}

% Not much to change here either!

\begin{figure*}
  \centering
  \includegraphics[scale=0.85]{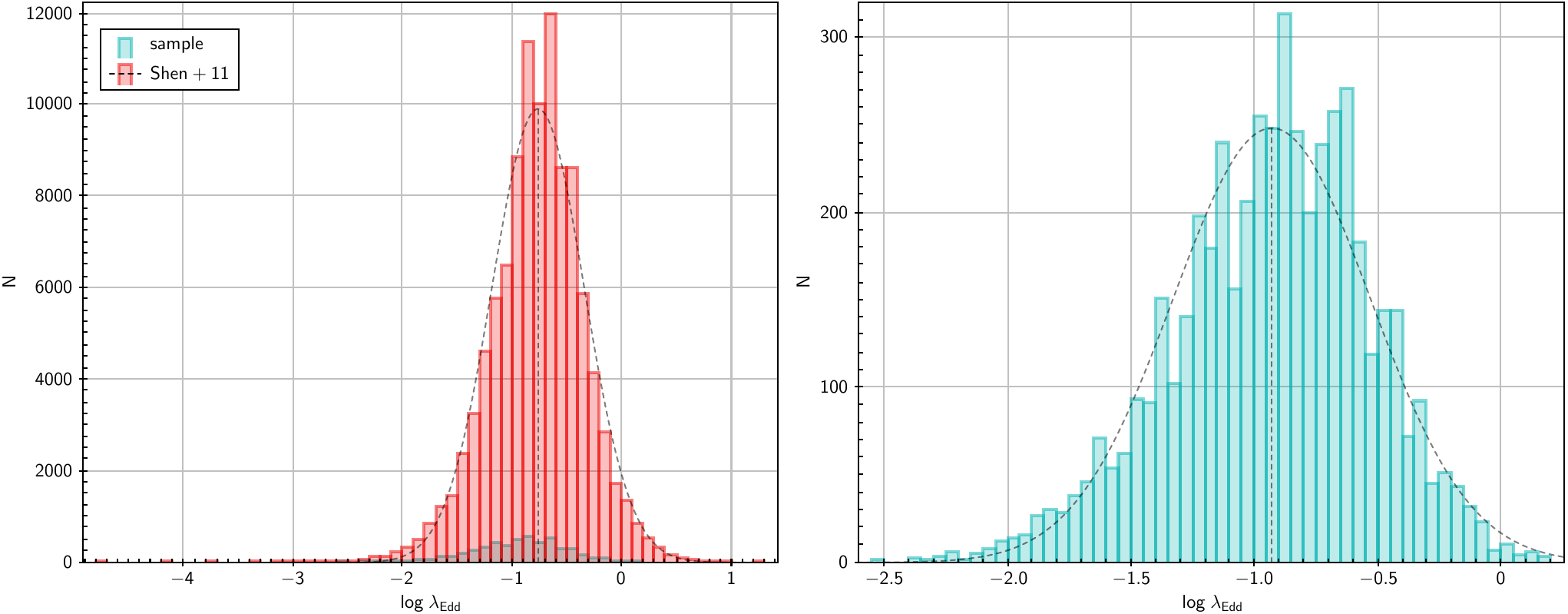}
  \caption{(a) Distribution of the $\lambda_\mathrm{{Edd}}$ from \citep{shen11} (in red). The error-limited sample (in turquoise) is shown underneath the whole distribution. The fitted gaussian has $\bar{x}$ = -0.769176, $\sigma$ = 0.42134285. (b) An enlarged version of the error-limited sample. The fitted gaussian has $\bar{x}$ = -0.93127483, $\sigma$ = 0.40046898.\label{fig:edd}}
  \label{fig:edd_shen}
\end{figure*}

We aim at reproducing the observed quasar optical plane from the physically motivated model of an accretion disk with a corona illuminating the BLR. Computations for a single quasar required assuming only the black hole mass and accretion rate (or alternatively, disk maximum temperature and the Eddington ratio), cloud number density, cloud column density, turbulent velocity and metallicity. For a whole quasar population, we fix the black hole mass range at $10^6 M_{\odot}$ to $3 \times 10^9 \odot$, and bolometric luminosities range was chosen from 0.01 to 1.0. 
The global model of the optical plane is then set by the remaining parameters.

The choice of the Eddington ratio range may be particularly important so we checked the distribution of this parameter in the \citet{shen11} quasar sample. The mean value of the Eddington ratio in the full \citet{sh14} catalog and in the sub-sample is close to 0.1, as demonstrated in Fig.~\ref{fig:edd_shen}. The mean value of the black hole mass in \citep{shen11} catalog is $\log M = 8.4$ if we limit ourselves to objects with measured H$\beta$, i.e. for redshift $z$ below $\sim 0.75$, (Panda et al. 2018, in preparation), and the corresponding value of $\log \mathrm{T_{max}}$ is about 4.80.

Since modelling Fe II emission caused significant problems in the past we first compare the model predictions for the mean values of the black hole mass and Eddington ratio in the sample with values measured by \citet{shen11} and \citet{sh14}. 

The mean and the median in the whole \citet{shen11} catalog for objects with measured $\mathrm{R_{Fe}}$ are  0.97 and 0.70, respectively, but if we limit ourselves only to the high quality sub-sample with low measurement errors, then the corresponding values drop to 0.64 and 0.38 \citep{sniegowska18}. Our value from the model, for the mean quasar parameters, is in the range 0.2 - 0.5, depending on the local density of the clouds. For the median Eddington ratio in the sample, 0.1, low density clouds predict too faint Fe II, but for the density $10^{12}$ cm$^{-3}$, and $10^{24}$ cm$^{-2}$ for the column density, the obtained value is $\sim 0.3$, if no turbulence and only Solar metallicity is assumed. If we allow for a turbulence of order of 10 - 20 km s$^{-1}$ then $\mathrm{R_{Fe}}$ rises to 0.32 - 0.35. Increasing the Fe II abundance by a factor of 3, combined with the turbulent velocity 10 km s$^{-1}$ increase $\mathrm{R_{Fe}}$  up to 0.9. Thus, even with very moderate increase of metallicity we reproduce well the mean value of $\mathrm{R_{Fe}}$ in the optical plane. This  itself is interesting since we have only few arbitrary parameters ($\mathrm{T_{max}}$, $\lambda_\mathrm{{Edd}}$,  $\mathrm{n_H}$, and $\mathrm{N_H}$), and in this case two of them are actually fixed by observations.

Thus, on average, we do not need any additional strong turbulent heating to explain the typical $\mathrm{R_{Fe}}$ ratios. Simple radiative reprocessing works well which is consistent with the possibility of the reverberation mapping of H$\beta$ and Fe II. Thus our relatively simple model works well for the average quasar parameters. We only need rather large densities and column densities, $10^{12}$ cm$^{-3}$, and $10^{24}$ cm$^{-2}$.

\begin{figure*}
  \centering
\includegraphics[height=9cm,width=13cm]{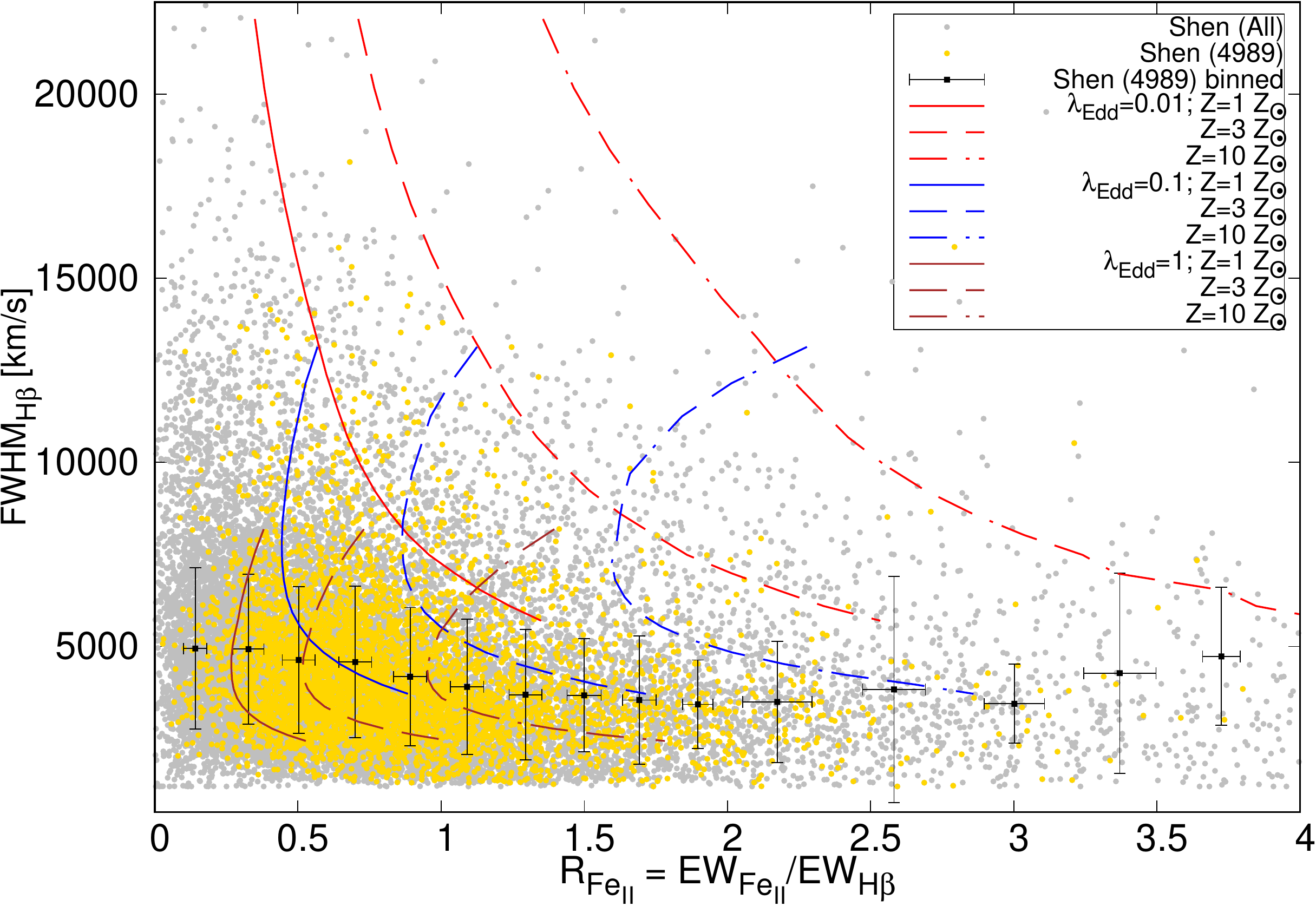}
% \vspace{0.5cm}
%  \includegraphics[scale=0.5, angle=0]{She\mathrm{n_H}o.pdf}
  \caption{Quasar optical plane: Comparison of $\mathrm{\mathrm{v_{FWHM}}}$ - $\mathrm{R_{Fe\;II}}$ obtained from the photoionisation simulations with observations \citep{shen11}. We consider 3 cases of $\lambda_\mathrm{{Edd}}$ = 0.01, 0.1 and 1.0, three values of the Fe II abundance: solar, 3 times solar, and 10 times solar, for a fixed cloud density  $\mathrm{n_H} = 10^{12} \;$ cm$^{-3}$, column density $10^{24}$ cm$^{-2}$, and turbulent velocity 10 km s$^{-1}$. The complete sample from the observations (105,783 objects) are shown in grey. The error-limited sample (4989 objects) are shown in gold. The black points with errorbars represent the average for the selected bins based on the $\mathrm{R_{Fe\;II}}$ values for the error-limited sample. \label{fig:optical_plane}}
  \end{figure*}

With this knowledge, we choose large cloud density and column density, a broad range of metallicity, and we calculated the results for the quasar sample. We overplotted the expected trends on the observational optical plane of EV1. We use the values $\mathrm{R_{Fe}}$ and FWHM as obtained from the computations for the range of densities and Eddington ratios (see Fig.~\ref{fig:optical_plane}). The model well covers the optical plane occupied by the data points.  Some of the trends are consistent with expectations. Large values of the FWHM of H$\beta$ correspond to lower values of the Eddington ratio. This simply reflects the relation between the accretion rate, black hole mass, line width and the SED peak position. Curves for solar abundance cover the region occupied by quasar majority but the high $\mathrm{R_{Fe}}$ (above 1) appear only from the objects with assumed super-Solar Fe II abundance.

The effective rise in the Fe II strength with increasing metallicity occurs with a caveat - the models shift rightwards when compared with the optical plane coverage, and thus, only predict the trend for objects with very high FWHM (see Figure \ref{fig:optical_plane}). Thus, it is not just enough to increase the metallicity to very high values to cover the high Fe II emitters but needs to be coupled with other parameters in such a way to cover the region of large $\mathrm{R_{Fe}}$ and small FWHM. On the other hand, this region is mostly populated by objects with low data quality (grey dots in Fig.~\ref{fig:optical_plane}) and high quality data point there (yellow dots) are rare.

\begin{figure*}
  \centering
  \includegraphics[scale=0.6, angle=270]{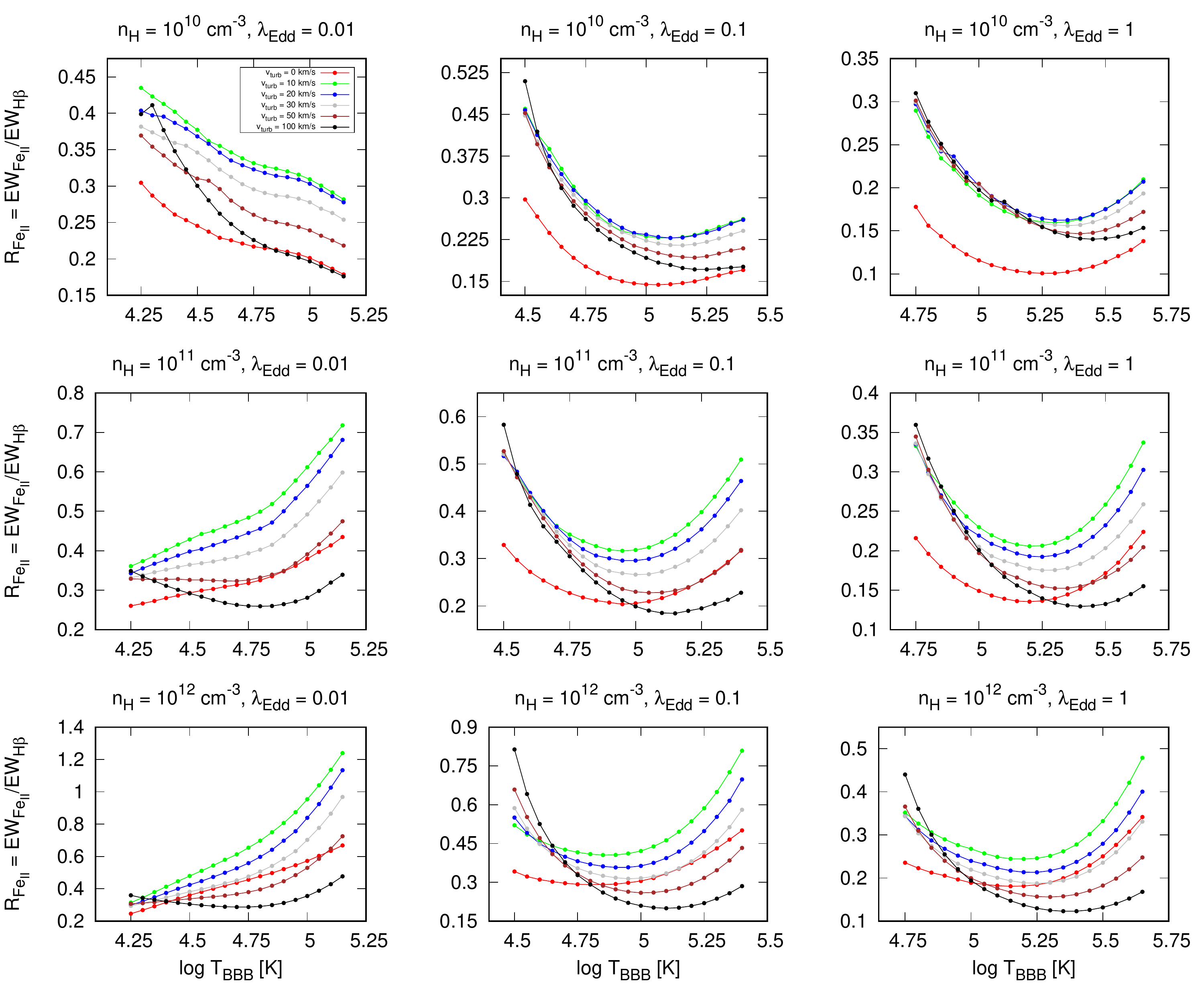}
  \caption{(From top to bottom:) The influence of the turbulent velocity on the Fe II production for varying cloud density ($10^{10}$ - $10^{12}$ cm$^{-3}$). (From left to right:) The influence of the turbulent velocity on the Fe II production for varying Eddington ratios ($\lambda_{\mathrm{Edd}}$: 0.01 - 1). Assumed abundance: solar. \label{fig:turb2}}
\end{figure*}

The dependence on the Eddington ratio is as not simple as postulated by \citet{bor02}. The Eddington ratio does not change monotonically along the main sequence, it actually change rather perpendicularly to it, and the impression of the overall increase of the Eddington ratio comes from the fact that most strong Fe II emitters have narrow lines, and high Eddington ratio objects in our model concentrate toward the bottom of the diagram. Therefore, the Eddington ratio cannot be identified as a single driver of the quasar main sequence in the optical plane.

We thus test the dependence of the Fe II emissivity on the SED shape since this is another potentially promising driver of the quasar main sequence. With this aim we show the dependence of the  $\mathrm{R_{Fe}}$ on the disk maximum temperature. If $\mathrm{T_{max}}$ is actually the expected driver, this dependence should be monotonic. The results of the computations for a range of Eddington ratios, densities and turbulent velocities is shown in (see Fig.~\ref{fig:turb2}). We see that the dependence in general is non-monotonic, particularly for moderate and high Eddington ratio. At the lowest Eddington ratios it is almost monotonic but the direction of the change depends critically on the local density of the clouds.

\begin{figure*}
  \centering
  \includegraphics[scale=0.7]{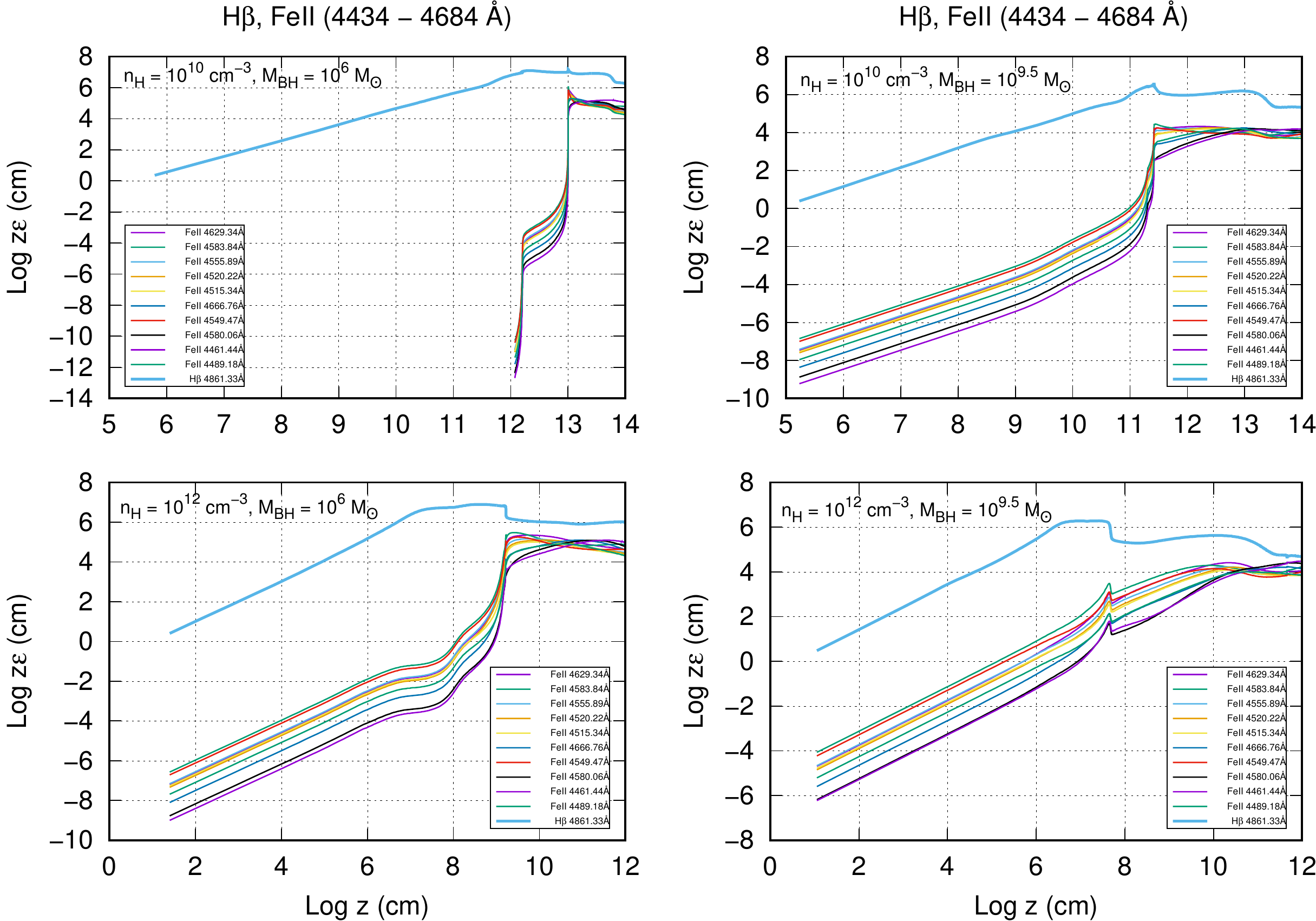}
  \caption{The emissivity profile for two clouds as function of the cloud depth measured from the illuminated
    surface. Since the plot is logarithmic we plot depth times emissivity to show clearly location of the emission peak. For low temperature cloud the hydrogen ionization front is visible and Fe II emission dominated at the dark side of the cloud. \label{fig:structure}}
\end{figure*}

The decrease of the Fe II emissivity with the rise of $\mathrm{T_{max}}$ happens since, for a fixed Eddington ratio, the distance to the BLR rises more slowly than the bolometric luminosity of the accretion disk, and the incident flux increases. In addition, the contribution of the hard X-ray power law also decreases, contributing to this overall trend. The cloud becomes more ionized, and the hydrogen ionization front visible for cold clouds disappears. To illustrate this phenomenon we plot two examples of the emissivity profiles of H$\beta$ and 10 strongest Fe II transitions (see Fig.~\ref{fig:structure}). These plots also show why large cloud density is required for efficient production of the Fe II: less dense clouds are more highly ionized, and Fe II production is less efficient. Also the role of the column density is clear: iron emission forms predominantly inside and at the back (dark side) of the cloud. We did not consider higher column densities than $10^{24}$ cm$^{-2}$.

%The model, however, shows a very strong dependence of the calculated $\mathrm{R_{Fe}}$ on the adopted cloud density. The local density $\log n = 12$ leads to much more efficient formation of the iron lines. This, combined with low or moderate values of $\lambda_\mathrm{{Edd}}$, allows values of $\mathrm{R_{Fe}}$ above 1.0. Lower values of the density, on the other hand, allow lower values of $\mathrm{R_{Fe}}$ to be obtained. Thus, the local cloud density seems to be a key parameter in shaping the overall trend of the quasar main sequence. In the presented model the cloud density is not constrained, and not coupled to the other key parameters of an active nucleus like its Eddington ratio and the mass. 

\section{Discussion}
\label{sec:discussion}

% Just made the Eddington ratio symbol definition earlier here and rewrote a couple of sentences

We constructed a simple but realistic model representing the physics of the line formation in AGN with the aim to reproduce the quasar main sequence. Our model is parameterized by the disk maximum temperature, $\mathrm{T_{max}}$, $\lambda_\mathrm{{Edd}}$ and the local cloud density ($n$). These values allow the building of broad band SEDs of AGNs, the determination of BLR locations, the emissivity of optical Fe II lines and H$\beta$ from CLOUDY code, and the calculation of H$\beta$ line widths. Allowing for a realistic range of values, we analyzed the coverage of the optical FWHM vs. $\mathrm{R_{Fe}}$ plane by the model and the observational data from the quasar sample. The mean values from the model and the data are consistent under the assumption of Solar metallicity, if we assume cloud density $10^{12}$ cm$^{-3}$, column density $10^{24}$ cm$^{-2}$, and turbulent velocity 10 - 20 km s$^{-1}$. For the same parameters, we represent well the whole optical plane if we allow for enhanced metallicity in some of the clouds. This is necessary to explain the extreme Fe II emitters.

It is very interesting that we are able to reproduce the Fe II emissivity under the assumption of purely radiative processes in the clouds. In the past, the need for additional collisional heating to achieve efficient Fe II production was postulated \citep{jol87,2001sac..conf..117N,baldwin2004}. Radiative driving of Fe II emission is strongly supported by the measured time delays of Fe II with respect to the continuum \citep{hu08}. It is also important to note that we assume a single production zone for Fe II and H$\beta$, and the clouds responsible for the line emission have universal density and column density. Clearly, accurate modelling of the line rations in specific objects require more complex approach, with the range of radii and densities \citep[e.g.][]{moloney2014,costantini2016}, and such modelling is frequently done within the Locally Optimized Cloud (LOC) model \citep{baldwin1995}. However, apparently the statistical distribution concentrates around the values based on simple and direct estimates. The universal column density in our model is roughly consistent with  predictions of the thermal instability in the irradiated medium introduced by \citet{krolik1981}, and later discussed in a number of papers \citep[e.g.][]{begelman1983,rozanska1996,krolik2001,krongold2003,danehkar2018}. The local density in turn is roughly consistent with the radiation pressure confinement of the BLR clouds nicely discussed by \citet{baskin18} in the Introdution to their paper. The self-consistency of the picture indicated that we are now making a considerable progress in the understanding of the BLR physics, going beyond the predominantly parametric models.

Some of the parameters cannot be derived yet from the basic constraints, like turbulent velocity, and metallicity. Also geometrical setup is not yet firmly set from the first principles although a major steps forward has been done with the development of the dust-based model of BLR formation \citep{ch11,czerny2015,czerny2017,baskin18}. We thus discuss below the particular aspects connected with these parameters.

\subsection{turbulence}

The emitting medium is most likely turbulent, and the turbulence decreases the optical depth of the clouds for lines. Fe II emissivity is quite sensitive to this value \citep[e.g.][]{bruhweiler08}. Therefore, we performed tests of the influence of the turbulence velocity of the calculated $\mathrm{R_{Fe}}$,  varying it between 0 and 100 km s$^{-1}$.  Overall, the Fe II emissivity increases but the trend is not monotonic. The emissivity rises with the rise of the turbulent velocity up to 20 km s$^{-1}$ and further increase in the velocity leads to a decrease of Fe II production apart from the low temperature tail (see Fig.~\ref{fig:turb2}). Thus the turbulence in the range 10-20 km s$^{-1}$ is generally favored for more efficient Fe II production, at such values are actually favored in detailed fitting of specific objects \citep[e.g.][]{bruhweiler08,hryniewicz14,marziani13,sredzinska17}. 

%This affects the way how the theoretical points cover the optical plane. We give an example of such coverage in Fig.~\ref{fig:shen_turb}. for the cloud density $10^{12}$ cm$^{-3}$. Such a plot still cannot reproduce the most extreme values of $\mathrm{R_{Fe}}$ but covers well most of the range of yellow dots marking the high quality data fits.

%\begin{figure*}
%  \centering
%    \includegraphics[height=7cm,width=18cm]{fig10_merged.pdf}
%  \caption{The coverage of the optical plane by the model as in Fig.~\ref{fig:optical_plane} but assuming the cloud density $10^{12}$ cm$^{-3}$ and the turbulent velocity 10 km s$^{-1}$(left panel) and 20 km s$^{-1}$ (right panel). %\label{fig:shen_turb}}
%\end{figure*}

\subsection{constant pressure clouds}

In the computations above we assumed a constant density model, traditionally adopted in the computations of the BLR clouds (\citealt{1977ApJ...218...20D}). However, physically it is not justified, particularly for such thick clouds since they are irradiated from the side exposed to the radiation flux from the central region, while the other side of the cloud is relatively cold. Such a cloud cannot be in hydrostatic equilibrium and would be rapidly destroyed. Therefore, more appropriate description of the cloud structure is to assume a constant pressure throughout the cloud, as discussed for Narrow Line Region clouds \citep[e.g.][]{davidson72}, warm absorber \citep[e.g.][]{rozanska06,adhikari15}, and BLR clouds \citep[e.g.][]{baskin14}. The effect is particularly strong for low local density clouds, but for high density clouds at BLR distances the compression is relatively less important, by a factor of a few \citep{adhikari18}. Nevertheless, the effect may be noticeable. We thus calculated an example of the constant pressure cloud corresponding to the most typical values for the observational sample: $\log \mathrm{T_{max}} = 4.8$, $\lambda_\mathrm{{Edd}} = 0.1 $, $\log n = 12$ at the cloud illuminated surface, and $\log \mathrm{N_H} = 24$. The results are shown in Table~\ref{tab:const_density}. The value of $\mathrm{R_{Fe}}$ calculated for such cloud was somewhat higher than for constant density cloud, if the turbulent velocity was neglected, and the effect decreased with the rise of the turbulence velocity. Thus, for such dense clouds, constant density and constant pressure models give very similar results.  

\begin{table}
\centering
\caption{Fe II strength comparison between constant density (CD) and constant pressure (CP) single cloud with microturbulence effect($\log \mathrm{T_{max}}$ = 5, $\log M/M_{\odot}$ = 8.5787)}
\label{tab:const_density}
\begin{tabular}{llll}
\hline\hline
$\mathrm{v_{turb}}$ (km/s) & $\mathrm{R_{Fe\;II}}\;$(CD) & $\mathrm{R_{Fe\;II}}\;$(CP) & $\Delta \mathrm{R_{Fe\;II}}$\footnote{$\Delta \mathrm{R_{Fe\;II}}$ = $\mathrm{R_{Fe\;II}}\;$(CP) - $\mathrm{R_{Fe\;II}}\;$(CD)} \\
\hline
0                 & 0.305          & 0.350          & 0.044            \\
10                & 0.421           & 0.434          & 0.013            \\
20                & 0.364          & 0.355          & -0.008      \\

%0                 & 0.3054          & 0.3496          & 0.0442            \\
%10                & 0.421           & 0.4337          & 0.0127            \\
%20                & 0.3635          & 0.3553          & -0.0082      \\
\hline    
\end{tabular}
\end{table} 

\subsection{closed geometry and enhanced contribution of the cloud dark sides}

Clouds are irradiated from one side, and the dark side of the clouds have a different proportion in H$\beta$ and Fe II emission. In the computations shown throughout this paper we used a plane parallel geometry, and the line intensity was calculated from the \textit{Intrinsic line intensities} section of the main CLOUDY output. These include the combined emission from the dark and the bright side of the cloud. This approach is 
justified if an observer is not highly inclined, clouds do not shield each other, the reprocessing of the emission of one cloud by the other cloud is negligible, and the geometrical thickness of the BLR is relatively small. With these approximations, we see on average the same total illuminated and dark surfaces of all clouds.
However, if any of these assumptions is violated, the obtained $\mathrm{R_{Fe}}$ ratio will be different. Such extreme set-ups might be responsible for the extreme Fe II emitters which were recovered in Sect.~\ref{sec:results} only for super-Solar metallicity.

Thus another possibility is that the abundances are always solar but the BLR is so geometrically thick that covers most of the quasar sky, and the number of clouds so large that the reprocessing is important. To check this option we calculated one cloud model using the closed geometry. We assumed the same cloud parameters as in Sect.~\ref{sec:model}. In this case the increase of the $\mathrm{R_{Fe}}$ is not very large since two effect counter-act. One is that we now see basically the dark side of the clouds, but the other is that multiple scattering increases the local incident flux and the cloud ionization. In addition, in order not to heavily absorb the continuum, we need a gap in the cloud distribution just along the line of sight to the innermost part of the accretion disk.

The second possibility is similar to the one above but with the cloud number not as high so the cross-illumination of clouds can be neglected. Again, in this case we see more of the dark sides of the clouds than of the bright sides. Such a picture has been used by (\citet{2009ApJ...707L..82F}) where the Fe II emission has been modelled as coming from infalling clouds. In this case we calculated separately the $\mathrm{R_{Fe}}$ values for the bright side and for the dark side of the cloud, by using plane parallel approximation but with the \textit{sphere} command to store just the outward line emission, and the inward emission has been calculated as a difference between total (intrinsic) and outward line flux:
\begin{equation}
\mathrm{\mathrm{R_{Fe\;II}} (dark) = \frac{Fe\;II_{Intrinsic} - Fe\;II_{Bright}}{H\beta_{Intrinsic} - H\beta_{Bright}}. }
 \label{eq:dark}
\end{equation}
The results are given in Table \ref{tab:open_closed}.
In the case of the same cloud ($\log \mathrm{T} = 5, \log \mathrm{n} = 12, \log \mathrm{N_H}$ = 24), we obtain  $\mathrm{R_{Fe}} =0.2883 $ for the bright side and  $\mathrm{R_{Fe}} = 0.2782$ for the dark side. Thus, if the clouds cover the nucleus densely we do not see an enhancement in the Fe II production since the inter-cloud scattering increases the overall ionization. The situation is different if we calculate emission from the dark sides of the clouds in a standard plane geometry, when no such inter-cloud scattering is present.
In such a geometry, the Fe II emission from the dark side of the cloud measured with respect to $H\beta$ is enhanced by a factor of 6. So in rare cases, when we see predominantly the dark sides of the clouds we can reproduce $\mathrm{R_{Fe}} $ values as high as a few, required to explain the extreme data points without postulating super-Solar metallicity.

\begin{table*}
\centering
\caption{Open vs closed geometry - contribution from the dark side of the cloud}
\label{tab:open_closed}
\begin{tabular}{cccccccc}
\hline\hline
$\mathrm{\mathrm{R_{Fe\;II}}}$ & $ \mathrm{Intrinsic}$ & $\mathrm{Bright}$ & $\mathrm{H\beta\;(Int.)} $\footnote{Int. = Intrinsic; Bri. = Bright; values are integrated intensities in $\rm{erg\;cm^{−2}\;s^{−1}}$.} & $\mathrm{H\beta \;(Bri.)} $ & $\mathrm{Fe\;II\;(Int.)}$ & $\mathrm{Fe\;II\;(Bri.)}$ & $\mathrm{Dark}$\footnote{see Eq.\ref{eq:dark}} \\
\hline
open                 & 0.143          & 0.133 & 10$^{44.016}$   & 10$^{44.010}$ & 10$^{43.173}$ & 10$^{43.133}$ & 0.921            \\
closed                & 0.285           & 0.288 & 10$^{44.142}$   & 10$^{43.972}$ & 10$^{43.597}$ & 10$^{43.432}$ & 0.278            \\

%open                 & 0.1434          & 0.1329 & 10$^{44.016}$   & 10$^{44.010}$ & 10$^{43.173}$ & 10$^{43.133}$ & 0.9206            \\
%closed                & 0.2854           & 0.2883 & 10$^{44.142}$   & 10$^{43.972}$ & 10$^{43.597}$ & 10$^{43.432}$ & 0.2782            \\
\hline 
\end{tabular}
\end{table*} 

%The $\mathrm{R_{Fe\;II}}$ reported by CLOUDY for open and close geometry and parameters used in Fig. with structure are: open: 0.0323  closed: 0.0517
\subsection{metallicity}
\begin{figure*}
  \centering
    \includegraphics[height=6cm,width=17.5cm]{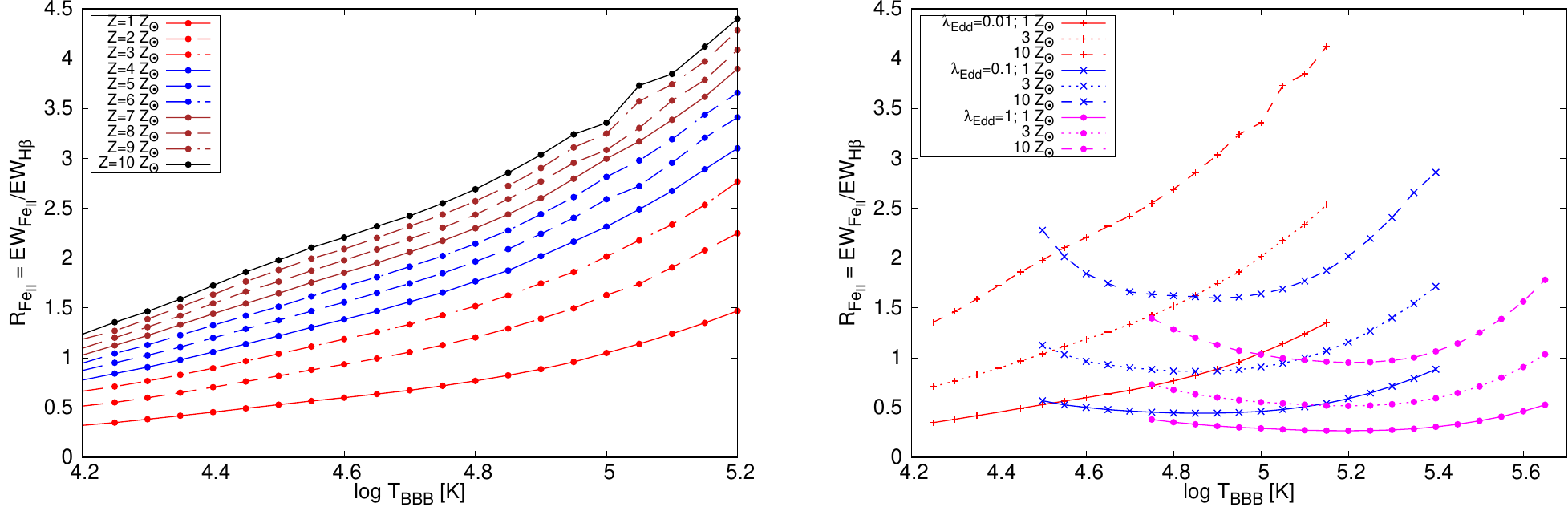}
  \caption{Effect of changing abundances starting from Solar (Z$_{\odot}$) to 10$\times$ Z$_{\odot}$. Here, we  have shown for the case with the maximum Fe II strength obtained - $\lambda_\mathrm{{Edd}} = 0.01,\;\log \mathrm{n_H} = 12\;[$cm$^{-3}],\;\mathrm{v_{turb}} = 10\;$km s$^{-1}$ (left panel). In the right panel the same effect is shown for three cases of abundances (Z/Z$_{\odot}$ =1, 3 and 10) with changing Eddington ratios ($\log \lambda_\mathrm{{Edd}}$ = 0.01, 0.1 and 1). \label{fig:abun}}
\end{figure*}

The proper coverage of the optical plane including the right corner occupied by the extreme Fe II emitters required allowing for a super-solar abundances, although the average quasar parameters were well represented without (see Sect.~\ref{sec:results}). The increase of the abundances simply enhances the Fe II emissivity, shifting the theoretical curves to the right in the optical plane (see Fig.~\ref{fig:optical_plane}). The same effect is seen if the parameter $\mathrm{R_{Fe}}$ is studied directly as a function of the maximum disk temperature. We show the corresponding plots in Fig.~\ref{fig:abun}.

%\begin{figure}
%  \centering
%    \includegraphics[scale=0.31]{abun_shen.pdf}
%  \caption{Coverage of the optical plane is shown for the same parameters as in Figure \ref{fig:abun} left panel. \label{fig:abun_shen}}
%\end{figure}

 Upon using the GASS model (at Z$_{\odot}$), we find an increase in the Fe II strength by 7 - 9 $\%$ in all the three cases of changing microturbulence compared to the default case for solar metallicity. The Fe II strength increases by 95 - 118 $\%$ and by 237-406 $\%$ when the metallicity (Z) is increased to 3Z$_{\odot}$ and 10Z$_{\odot}$, respectively (Table \ref{tab4}).  In this case only the most extreme values of $\mathrm{R_{Fe}}$ (above 4) are not accounted for and might require that we see predominantly dark sides of the clouds (see Figure \ref{fig:abun}). 

\subsection{extreme EV1 objects}

In \citet{sniegowska18}  27 objects were selected for study from the \citep{shen11} catalog, but after careful analysis only 6 objects were
confirmed as strong Fe II emitters. Three of these sources had broad H$\beta$ lines, with FWHM above 4500 km s$^{-1}$  the mean black hole mass $\log M = 8.9$, and the Eddington ratio about 0.01, while the other three had very narrow H$\beta$ (below 2100 km s$^{-1}$), mean black hole mass $\log M = 7.5$, and the Eddington ratio above 0.3. The first family of quasars is consistent with high expected values of $\mathrm{R_{Fe}}$ since the typical maximum temperature in this case is about 20 000 K. The second group has the temperatures of the order of $2 \times 10^5$ K, and from the model computations the expected values of $\mathrm{R_{Fe}}$ are low, particularly for high Eddington ratio sources. Model predicts only the further rise of $\mathrm{R_{Fe}}$ if the temperatures are well above $10^6$ K. This cannot happen within the frame of the black body representation of the Big Blue Bump.   

\subsection{effect of BLR size}
\begin{figure}
  \centering
    \includegraphics[height=6cm,width=8.5cm]{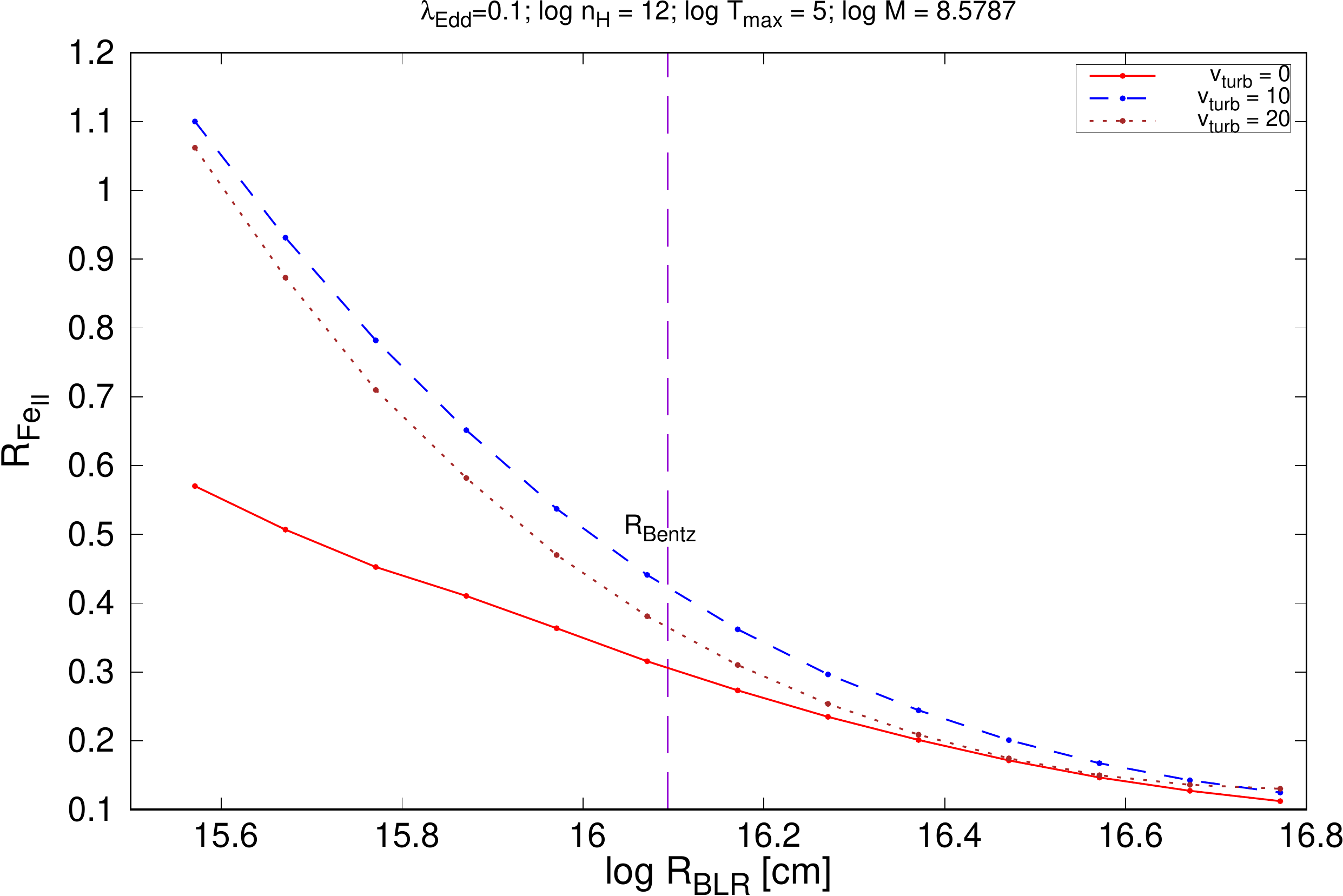}
  \caption{Effect of changing R$_\mathrm{{BLR}}$ on the Fe II strength: Changing the microturbulence ($\mathrm{v_{turb}}$). The vertical dashed line represents the radius value used from the \citet{bentz13}.\label{fig:radius}}
\end{figure}

The inner radius of the BLR cloud that has been used in this paper follows the \citet{bentz13}  (Sec. \ref{blr}). However, the BLR is actually extended. Here we test the effect of changing the radius from 0.3R$_{\mathrm{Bentz}}$ to 5R$_{\mathrm{Bentz}}$. The lower limit (0.3R$_{\mathrm{Bentz}}$) used has been set corresponding to maximum disk temperature ($\sim$2000 K) expected from the BLR model based on dust presence in the disk atmosphere (\citealt{2011A&A...525L...8C}). The upper limit is set assuming $\mathrm{R_{out} \sim 5R_{\mathrm{Bentz}}}$ comes from the dust reverberation studies of the torus \citep{koshida14}. The results for one such case is shown Figure \ref{fig:radius}. There is a monotonic behaviour of the Fe II strength with respect to changing BLR radius. In a single zone approximation Fe II emission is relatively more efficient if the BLR is located closer in, with all the other parameters fixed. This suggests that future studies should include the radial stratification of the BLR but it is not simple since the results would depend on the weighted emissivity as a function of radius. Additionally, continuum is variable and BLR responds to it after a delay. One needs to take present-day continuum luminosity and line width which traces radius corresponding to continuum luminosity from the past. Single epoch spectra, as in SDSS, do not trace this effect which introduces some bias.  Study of the full BLR structure is beyond the scope of the current work.

\subsection{viewing angle}

In our model we did not include the range of viewing angles towards the nucleus. As pointed out by many authors \citep{2002ChJAA...2..487Z,c06,sh14}, BLR is not flat. If the emission of Fe II and H$\beta$ comes roughly from the same region, as assumed in the current paper, the ratio $\mathrm{R_{Fe}}$ is not affected but the measured H$\beta$ width is expected to depend on the viewing angle, $i$. In type 1 sources this viewing angle is never very large, otherwise the torus shields the view of the nucleus. The frequently adopted range of the viewing angles is thus between 0 and 45 deg. \cite{Mar01} and others have tried to connect the observational plane with the source orientation and Eddington ratio. The line width, in turn, depends both on $\sin i$, and the turbulent (random) velocity field. Thus the FWHM can be affected by a factor of 2 for the BLR thickness of order of 0.3 (see Eq. 8 in \citealt{c06}), and similar effect was determined in the studies of the virial factor trends \citep{meija18}. Thus the vertical extension of the covered area can be increased by a factor less than 2 if this effect is included. On the other hand, we see from our modelling that the spread in the vertical direction is mostly caused by the coupling between the model parameters and the Fe II production efficiency. We intend to look into the effects of the orientation on the main sequence more carefully in our subsequent work.

%\begin{figure}[ht!]
%\plotone{mrk2.pdf}
%\caption{Polarization fraction in per cent with errors (with offset (Cikota et. al. 2017)) for the Mrk1044 zoomed in close to the emission line. The total intenstiy spectrum is overplotted to show the behaviour of the $P_L$ variation close to the H$\alpha$ emission line. On the X-axis is the observed frame wavelength. \label{fig:star}}
%\end{figure}

%% Note that the \setcounter and \renewcommand are needed here because
%% this example is using a mix of deluxetable and tabular.  Here the
%% deluxetable counters are set with \tablenum but the situation is a bit
%% more complex for tabular.  Use the first command to set the Table number
%% to ONE LESS than it should be.  The next command will auto increment it
%% to the desired number.
\subsection{Correlation between the Eddington ratios and the black hole masses}
\begin{figure*}
\begin{minipage}[c]{0.4\linewidth}
\includegraphics[height=6cm,width=8.5cm]{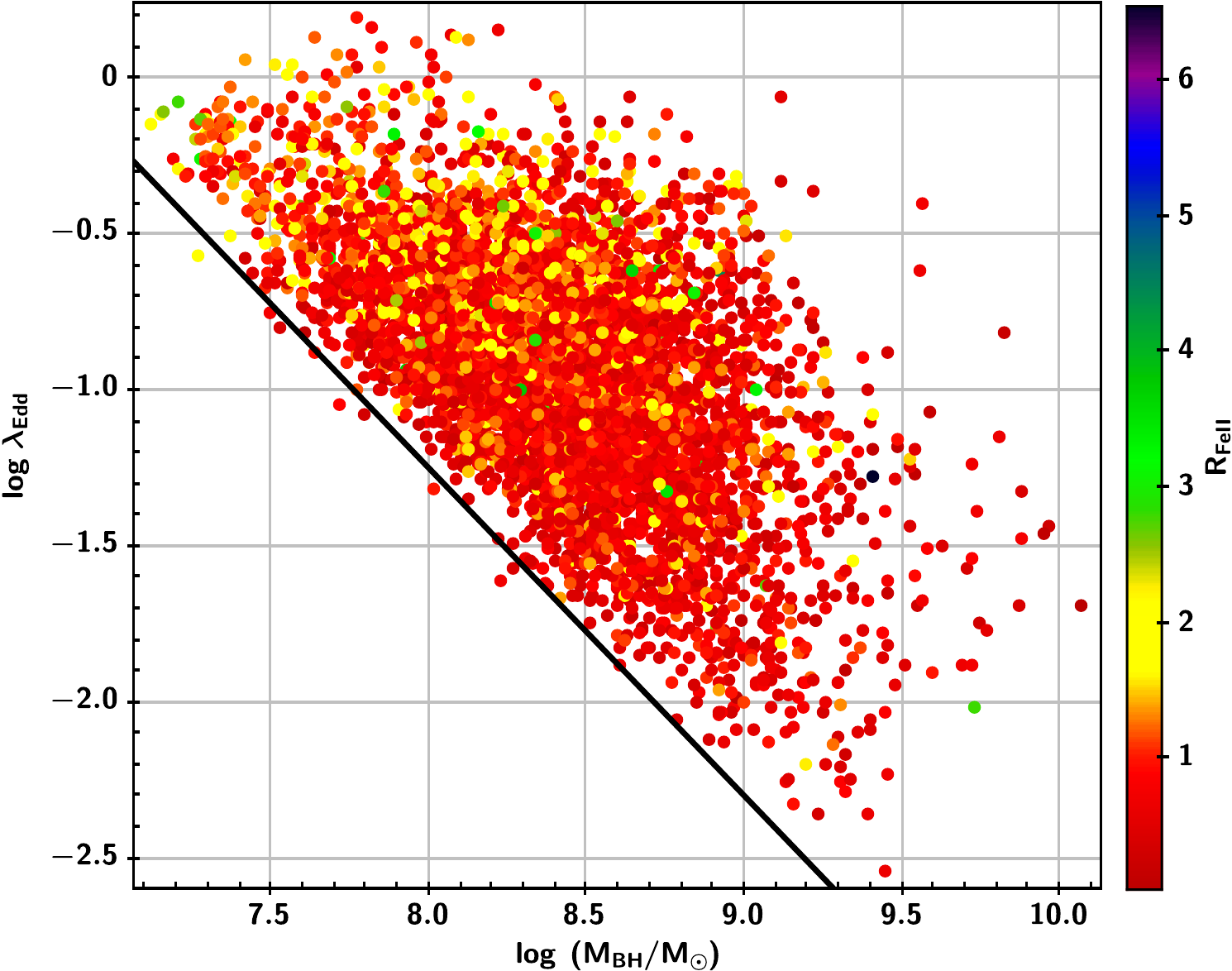}
\end{minipage}
\quad\quad\quad\quad\quad 
\begin{minipage}[c]{0.4\linewidth}
\includegraphics[height=6cm,width=8.5cm]{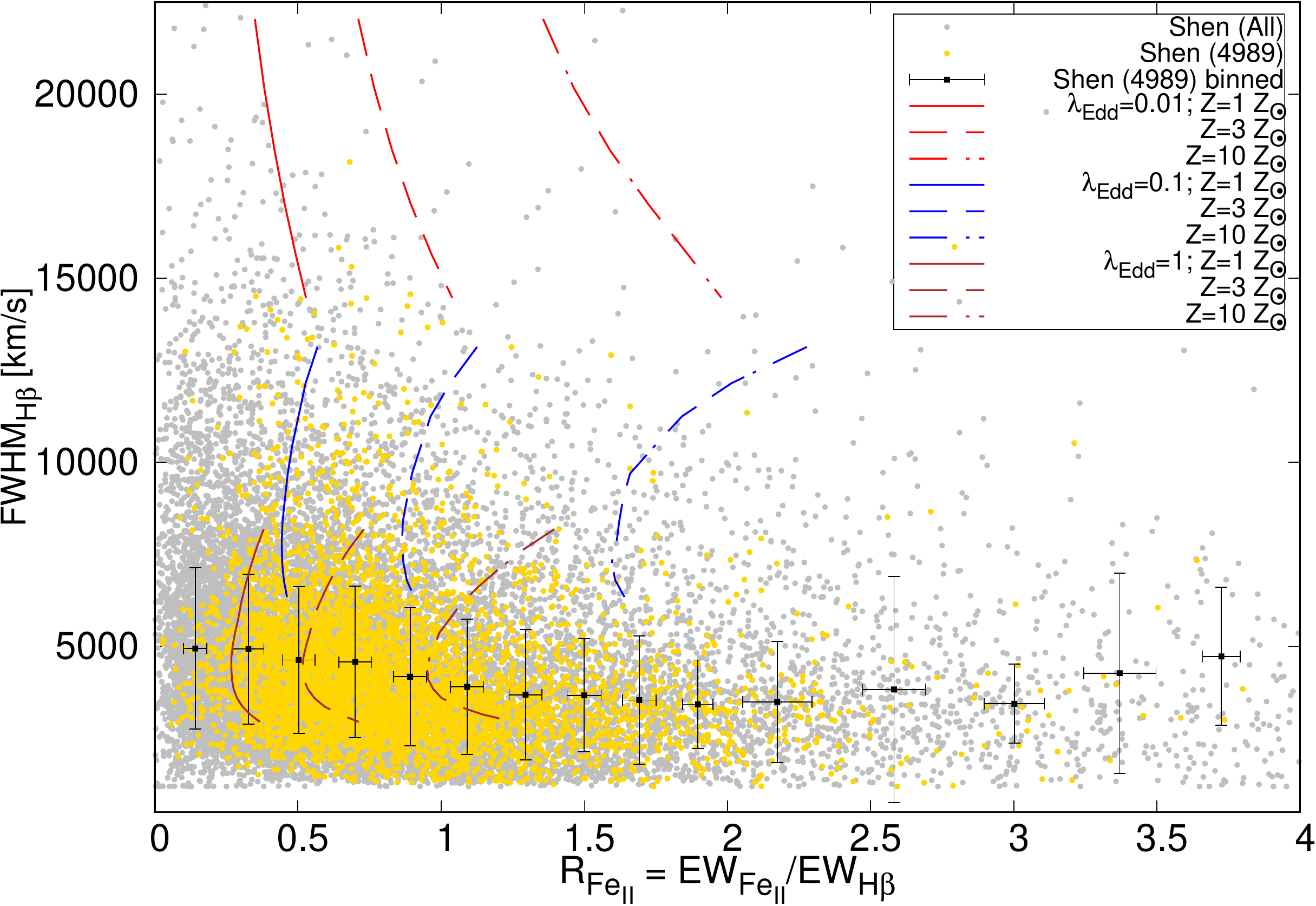}
\end{minipage}%
\caption{Left panel shows the $\mathrm{log \; \lambda_{Edd}- log\;M_{BH}}$  plot for the error-limited quasar sub-sample from the \cite{s11} (4989 objects). The auxilary axis shows the distribution of the FeII strength. The black solid line (see Equation \ref{Eq10}) is used to constrain the modelling of the quasar optical plane for real observations (right panel).}
\label{fig:ref_new}
\end{figure*}

In our approach we have not yet explicitly connected the Eddington ratios to the black hole masses in the observed sample. If we plot the quality-controlled sample of 4989 objects from the original \cite{s11},  the Eddington ratio and the black hole masses in current optical plane can be constrained with: 
\begin{equation}
\mathrm{log\;\lambda_{Edd}} = -1.05\;\mathrm{log\;M_{BH}} + 7.15;
\label{Eq10}
\end{equation}
 where $\mathrm{M_{BH}}$ is considered in  $\mathrm{M_{\odot}}$. This is shown in left panel of Figure \ref{fig:ref_new}. If the analysis to model the optical plane is to be contained within the limits of this boundary defined between the Eddington ratio and the black hole mass, then we cover a much lesser portion of the optical plane than before (ssee right panel of Figure \ref{fig:ref_new}). Still, with this additional constraint, we cover 4903 out of the 4989 objects i.e., over 98\% of the total sub-sample. Indeed, as shown in Figure \ref{fig:ref_new}, we require deeper observations to exploit the lower regime of the $\mathrm{log \; \lambda_{Edd}- log\;M_{BH}}$ in the context of the quasar optical plane.

\section{Conclusions}

% Just rewrote the third bullet point for clarity

We model the quasar main sequence using realistic description of the quasar broad band spectra, assuming the distance to the BLR as known from reverberation measurements and calculating the line emission using CLOUDY, version C17. We show that

\begin{itemize}
\item mean quasar parameters are well reproduced by our single zone constant density model and solar abundance, particularly if we take into account the turbulent velocity of order of 10 - 20 km s$^{-1}$
\item high density clouds ($n \sim 10^{12}$ cm$^{-3}$) allow good coverage of the optical plane; such densities are consistent with the radiation pressure confinement of the BLR clouds (\citealt{2014MNRAS.438..604B}, \citealt{2018ApJ...856...78A})
\item high values of $\mathrm{R_{Fe}}$ ($>$ 1) require higher abundance of iron and/or enhanced contribution from the cloud dark sides; in the second option very large solid angle of the BLR in these sources are required
\item the range of viewing angles is only partially responsible for the dispersion in the quasar main sequence; most of the dispersion comes from a range of black hole masses and accretion rates
\item the dependence of the $\mathrm{R_{Fe}}$ ratio neither on the Eddington ratio nor on the maximum disk temperature is not monotonic.
 \end{itemize}
%\vspace{5cm}
\begin{table*}
\begin{flushleft}
\caption{Effect of changing abundances on the obtained values of $\mathrm{\mathrm{R_{Fe\;II}}}$}
\label{tab4}
\resizebox{16cm}{1.05cm}{
\begin{tabular}{cccccccccc}
\hline
\hline
$\mathrm{\mathrm{v_{turb}}}$ & $\mathrm{\mathrm{R_{Fe\;II}}}$(def) & $\mathrm{\mathrm{R_{Fe\;II}}}(Z_{\odot}$) & ratio  & $\mathrm{\mathrm{R_{Fe\;II}}}(3Z_{\odot})$ & ratio & ratio & $\mathrm{\mathrm{R_{Fe\;II}}}(10Z_{\odot}$) & ratio  & ratio \\
$\mathrm{[km/s]}$& & &(col2/col3) & & (col5/col2)& (col5/col3) & & (col8/col2)& (col8/col3)\\
\hline
0 & 0.305 & 0.329 & 1.077 & 0.595 & 1.948 & 1.809 & 1.030 & 3.374 & 3.132 \\
10 & 0.421 & 0.462 & 1.096 & 0.907 & 2.153 & 1.964 & 1.634 & 3.882 & 3.540 \\
20 & 0.364 & 0.399 & 1.097 & 0.792 & 2.179 & 1.987 & 1.475 & 4.059 & 3.702 \\

%0 & 0.3054 & 0.329 & 1.077275704 & 0.595 & 1.9482645711 & 1.8085106383 & 1.0303 & 3.3736083824 & 3.1316109422 \\
%10 & 0.421 & 0.4616 & 1.0964370546 & 0.9066 & 2.1534441805 & 1.9640381282 & 1.6342 & 3.8817102138 & 3.5402946274 \\
%20 & 0.3635 & 0.3986 & 1.0965612105 & 0.792 & 2.1788170564 & 1.9869543402 & 1.4754 & 4.058872077 & 3.7014550928\\
\hline
\end{tabular}}
\end{flushleft}
\end{table*}

%\begin{figure}
%  \centering
%  \vspace{0.5cm}
%  \includegraphics[scale=0.5, angle =0]{rfe2_8to12.pdf}
%  \includegraphics[scale=0.3,angle =0]{fig5.pdf}
%  \caption{The same as Fig.~\ref{fig:R_Fe_basic} but for broader density range and only for $\lambda_\mathrm{{Edd}}=1$.  \label{fig:lower_density}}
%\end{figure}

\section*{Acknowledgements}
The project was partially supported by the Polish Funding Agency National Science Centre, project 2015/17/B/ST9/03436/ (OPUS 9). SP would like to acknowledge Dr. Abbas Askar for his assistance on using the computational cluster, Dr. Agata R{\'o}{\.z}a{\'n}ska and the high energy astrophysics group at CAMK for every engaging discussions we've had and will have in the near future. TPA would like to acknowledge the support from the Polish Funding Agency National Science Centre, project 2016/21/N/ST9/03311 (Preludium 9).
%\section*{Softwares}
\software {The results in this paper are mainly attributed to the photoionisation code CLOUDY (\citealt{f17}). The authors would like to acknowledge the use of data analysis software TOPCAT \citep{2005ASPC..347...29T}}.
\bibliographystyle{aasjournal}
\bibliography{eigenvector1}

\section*{appendix}

\begin{figure*}
  \centering
    \includegraphics[width=18cm,height=12cm]{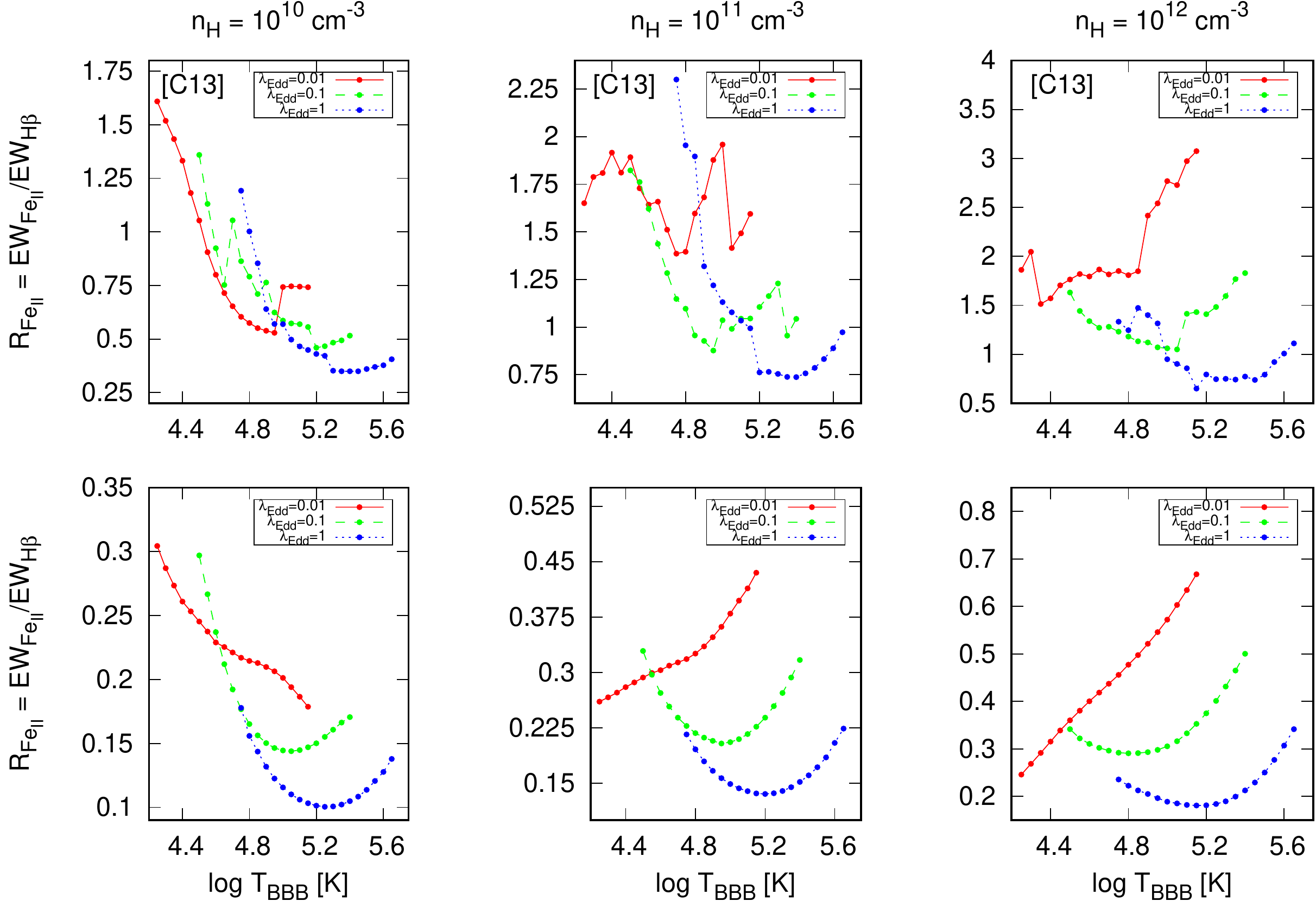}
  \caption{The comparison of the CLOUDY code version 13 with the CLOUDY code version 17. \label{fig:c1317}}
\end{figure*}

%\subsection{version of the CLOUDY code}
The atomic data available for radiative transfer are still under development and are far from satisfactory. This is reflected in the changes of the atomic data available in CLOUDY code.
At the beginning of the project we made some runs using the CLOUDY code version
 C13 \citep{f13}, but finally we changed to version C17 \citep{f17} and all results given here were obtained with the newest code. We always used the option \textit{species ``Fe+" levels=all} for most accurate computations of the Fe II emission, as stressed in CLOUDY's Hazy 1\footnote{https://www.nublado.org/} Manual. We show the comparison of the two code versions in Fig.~\ref{fig:c1317}. In general, the new code returns higher emissivity in H$\beta$ line and lower emissivity in the optical Fe II, so the net values of $\mathrm{R_{Fe}}$ are lower in the newest version of CLOUDY.

The use of the option \textit{species ``Fe+" levels=all} in version 17 is critical, since without it the Fe II emissivity is by $\sim 50$ \% higher than without this option. This option uses all the Fe II transitions, including \citet{verner99} Fe II model  while if the option is not on, only a simplified old model of \cite{1985ApJ...288...94W} is adopted to speed up the computations.

\end{document}